\documentclass[twocolumn,showpacs,amsmath,amssymb,prl,aps,floats]{revtex4}
\usepackage{bm}
\usepackage{graphicx}
\usepackage{times}
\newcommand{\beq}{\begin{equation}}
\newcommand{\eeq}{\end{equation}}

\newcommand{\bfb}{\mbox{\boldmath $b$}}

\newcommand{\bfh}{\mbox{\boldmath $h$}}

\newcommand{\bfu}{\mbox{\boldmath $u$}}
\newcommand{\bfv}{\mbox{\boldmath $v$}}
\newcommand{\bfx}{\mbox{\boldmath $x$}}
\newcommand{\bfB}{\mbox{\boldmath $B$}}
\newcommand{\bfH}{\mbox{\boldmath $H$}}

\newcommand{\bfR}{\mbox{\boldmath $R$}}
\newcommand{\bfX}{\mbox{\boldmath $X$}}
\newcommand{\bfxi}{\mbox{\boldmath $\xi$}}
\newcommand{\ex}{\mbox{{\boldmath $e$}}_{1}}
\newcommand{\ey}{\mbox{{\boldmath $e$}}_{2}}
\newcommand{\ez}{\mbox{{\boldmath $e$}}_{3}}
\newcommand{\bfemf}{\mbox{\boldmath ${\cal E}$}}
\newcommand{\bnabla}{\mbox{\boldmath $\nabla$}}
\newcommand{\cross}{\mbox{\boldmath $\times$}}
\newcommand{\cendot}{\mbox{\boldmath $\cdot\,$}}

\begin{document}


\title{Shear dynamo problem: Quasilinear kinematic theory}
\author{S. Sridhar}
\affiliation{Raman Research Institute, Sadashivanagar, Bangalore 560 080, India}
\email{ssridhar@rri.res.in}
\author{Kandaswamy Subramanian}
\affiliation{IUCAA, Post bag 4, Ganeshkhind, Pune 411 007, India}
\email{kandu@iucaa.ernet.in}

\date{\today}
\begin{abstract}
Large--scale dynamo action due to turbulence in the presence of a linear shear flow is studied. 
Our treatment is quasilinear and kinematic but is non perturbative in the shear strength.
We derive the integro--differential equation for the evolution of the mean magnetic field,
by systematic use of the shearing coordinate transformation and the Galilean invariance
of the linear shear flow. For non helical turbulence the time evolution of the cross--shear 
components of the mean field do not depend on any other components excepting themselves. 
This is valid for any Galilean--invariant velocity field, independent of its dynamics. 
Hence the shear--current assisted dynamo is essentially absent, although large--scale non helical 
dynamo action is not ruled out.
\end{abstract}

\pacs{47.27.W-, 47.65.Md, 52.30.Cv, 95.30.Qd}
\maketitle

Shear flows and turbulence are ubiquitous in astrophysical systems.
Recent work suggests that the presence of shear may open new 
pathways to the operation of large--scale dynamos 
\cite{BRRK08,Yousef08,PKB,RK03,Schek08}.
We present a theory of dynamo action in a shear flow of an 
incompressible fluid which has random 
velocity fluctuations due either to freely decaying turbulence 
or generated through external forcing.
Of particular interest is the case of non helical large--scale dynamo action
in shear flows. 
Several direct simulations show that large--scale fields
can grow from small seed fields under the combined action of non helical
turbulence and background shear flow \cite{BRRK08,Yousef08}. 
However, the interpretation of how such a dynamo works is not yet clear.
One possibility that has attracted much attention is the shear--current effect \cite{RK03},
in which extra components of the mean electromotive force (EMF) arise due to shear,
which couple components of the mean magnetic field parallel and perpendicular to
the shear flow. However there is no convergence yet on whether the sign
of the relevant coupling term is such as to obtain a dynamo;
some analytic calculations \cite{RS06,RK06} and numerical experiments
\cite{BRRK08} find that the sign of the shear--current term is unfavorable 
for dynamo action. Moreover, analytic calculations treat shear as a small perturbation. 
We are interested here in studying the shear dynamo without such a restriction. 

Our theory is `local' in character: In the lab frame we consider a background shear flow whose velocity is unidirectional (along the $X_2$ axis) and varies linearly in an orthogonal direction (the $X_1$ axis). The linear shear flow has a basic symmetry relating to measurements made by a special subset of all observers, who may be called comoving observers. This symmetry is the invariance of the equations with respect to a group of transformations that is a subgroup of the full Galilean group. It may be referred to as `shear--restricted Galilean invariance', or Galilean invariance (GI). We introduce and explore the consequences of GI velocity fluctuations; not only are these compatible with the underlying symmetry of the problem, but they are expected to arise naturally. This has profound consequences for dynamo action, because the transport coefficients that define the mean EMF become spatially homogeneous in spite of the shear flow. Systematic use of the shearing transformation allows us to develop a theory that is non perturbative in the strength of the background shear. However, we ignore the complications associated with nonlinear interactions, hence MHD turbulence and the small--scale dynamo; so our theory is  quasilinear in nature, equivalent to the `first order smoothing approximation' (FOSA). 

Let $(\ex,\ey,\ez)$ be the unit vectors of a Cartesian coordinate system in the lab frame,  
$\bfX = (X_1,X_2,X_3)$ the position vector, and $\tau$ the time. The fluid velocity is given 
by $(-2AX_1\ey + \bfv)$, where $A$ is the shear parameter and $\bfv(\bfX, \tau)$ is a randomly fluctuating velocity field which is incompressible $(\bnabla\cendot\bfv = 0)$ and has zero mean $(\left<\bfv\right> = {\bf 0})$. The magnetic field has a large--scale (mean field) component $\bfB(\bfX, \tau)$, and a fluctuating field, $\bfb$, with zero mean $(\left<\bfb\right> = {\bf 0})$. The evolution of the mean field is governed by
\beq
\left(\frac{\partial}{\partial\tau} - 2AX_1\frac{\partial}{\partial X_2}\right)\bfB + 2AB_1\ey = 
\bnabla\cross\bfemf + \eta\bnabla^2\bfB\label{meanindeqn}
\eeq
\noindent
where 
$\bfemf = \left<\bfv\cross\bfb\right>$ is the mean EMF. Our goal is to calculate $\bfemf$ in
terms of the statistical properties of the fluctuating velocity field, 
which we will do using quasilinear theory. This means solving the equation 
for $\bfb$ by dropping terms that are quadratic in the fluctuations. 
We also drop the resistive term, assuming that the
correlation times are small compared to the resistive timescale. So our theory is 
applicable when FOSA is valid \cite{dynam}. Then $\bfb$ obeys
\beq 
\left(\frac{\partial}{\partial\tau} - 2AX_1\frac{\partial}{\partial X_2}\right)\bfb \;+\; 2Ab_1\ey = 
\bnabla\cross\left(\bfv\cross\bfB\right) 
\label{flucindeqn}
\eeq

It proves convenient to exchange spatial inhomogeneity for temporal inhomogeneity, so we 
get rid of the $\left(X_1{\partial/\partial X_2}\right)$ term through a shearing transformation to 
new spacetime variables,
\beq
x_1 = X_1\,,\quad x_2 = X_2 + 2A\tau X_1\,,\quad x_3 = X_3\,,\quad t = \tau
\label{sheartr}
\eeq
\noindent
We also define new variables, $\bfH(\bfx, t) = \bfB(\bfX, \tau),\bfh(\bfx, t) = \bfb(\bfX, \tau)$ and
$\bfu(\bfx, t) = \bfv(\bfX, \tau)$, which are component--wise equal to the old variables. 

Then equation~(\ref{flucindeqn}) becomes
\begin{eqnarray}
\frac{\partial\bfh}{\partial t} + 2Ah_1\ey &=& \left(\bfH\cendot\frac{\partial}{\partial\bfx}
 + 2AtH_1\frac{\partial}{\partial x_2}\right)\bfu -\nonumber\\ 
&&- \left(\bfu\cendot\frac{\partial}{\partial\bfx} + 2Atu_1\frac{\partial}{\partial x_2}\right)\bfH
\label{newvareqn}
\end{eqnarray}
\noindent
Not only do sheared coordinates get rid of spatial inhomogeneity, but in quasilinear theory the 
evolution equation~(\ref{newvareqn}) does not contain spatial derivatives  of $\bfh(\bfx, t)$. 
The equations for $h_1$ and $h_3$ can be integrated directly. The $h_1$ so obtained
can be substituted in the equation for $h_2\,$: there occur double--time integrals which can
be manipulated to give expressions with only single--time integrals, by interchanging the order of the 
integrals. Then the particular solution for $\bfh(\bfx, t)$ is given in component form by
\begin{eqnarray}
h_m &=& \int_0^t dt'\left[u'_{ml} - 2A(t-t')\delta_{m2}u'_{1l}\right]\left[H'_l + 2At'\delta_{l2}H'_1\right]
\nonumber\\
&-& \int_0^t dt'\left[u'_l + 2At'\delta_{l2}u'_1\right]\left[H'_{ml} - 2A(t-t')\delta_{m2}H'_{1l}\right]
\nonumber\\
&&\label{hsoln}
\end{eqnarray}
\noindent
where primes  denote evaluation at spacetime point $(\bfx, t')$. We have also used notation $u_{ml} = (\partial u_m/\partial x_l)$ and $H_{ml} = (\partial H_m/\partial x_l)$.

The expression in equation~(\ref{hsoln}) for $\bfh$ should be substituted in $\bfemf = \left<\bfv\cross\bfb\right> = \left<\bfu\cross\bfh\right>$. Following standard procedure, we allow $\left<\;\;\right>$ to act only on the velocity variables but not the mean field; symbolically, it is assumed that $\left<\bfu\bfu\bfH\right> = \left<\bfu\bfu\right>\bfH$. After interchanging the dummy indices $(l,m)$ in the last term, we find that the mean EMF is  
\begin{eqnarray}
{\cal E}_i = \int_0^t dt'\left[\widehat{\alpha}_{il} - 2A(t-t')\widehat{\beta}_{il}
\right]\left[H'_l + 2At'\delta_{l2}\,H'_1\right] \;-\; &&\nonumber\\
\int_0^t dt'\left[\,\widehat{\eta}_{iml} + 2At'\delta_{m2}\,\widehat{\eta}_{i1l}\right]\left[H'_{lm} - 2A(t-t')\delta_{l2}\,H'_{1m}\right]
\label{emf}&&
\end{eqnarray}
\noindent
where the transport coefficients $(\widehat{\alpha}\,,\widehat{\beta}\,,\widehat{\eta}\,)$ are 
defined in terms of the $\bfu\bfu$ velocity correlators by
\begin{eqnarray}
\widehat{\alpha}_{il}(\bfx, t, t') &\;=\;& \epsilon_{ijm}\left<u_j(\bfx, t)\,u_{ml}(\bfx, t')\right>\nonumber\\[1ex]
\widehat{\beta}_{il}(\bfx, t, t') &\;=\;& \epsilon_{ij2}\left<u_j(\bfx, t)\,u_{1l}(\bfx, t')\right>\nonumber\\[1ex]
\widehat{\eta}_{iml}(\bfx, t, t') &\;=\;& \epsilon_{ijl}\left<u_j(\bfx, t)\,u_m(\bfx, t')\right>
\label{trcoeffs}
\end{eqnarray}
\noindent
It is physically more transparent to consider velocity statistics in terms of the $\bfv\bfv$ velocity correlators, because this is referred to the lab frame, instead of the sheared coordinates. By definition,
\beq
u_m(\bfx, t) \;=\; v_m(\bfX(\bfx,t), t)
\label{uvtr}
\eeq
\noindent
where $\bfX(\bfx,t) = \left(x_1, x_2 - 2Atx_1, x_3\right)$ is the inverse of the shearing transformation given in equation~(\ref{sheartr}). The velocity gradient $u_{ml}$ is
\beq
u_{ml} = \left(\frac{\partial}{\partial X_l} - 2A\tau\,\delta_{l1}\,\frac{\partial}{\partial X_2}\right)\,v_m
= v_{ml} - 2A\tau\,\delta_{l1}\,v_{m2}
\label{uvgrad}
\eeq
\noindent
where $v_{ml} = (\partial v_m/\partial X_l)$. Using equations~(\ref{uvtr}) and (\ref{uvgrad}) in (\ref{trcoeffs}), 
\begin{eqnarray}
\widehat{\alpha}_{il}(\bfx, t, t') &=& \epsilon_{ijm}\left[\left<v_j(\bfX, t)\,v_{ml}(\bfX', t')\right>\right.
\nonumber\\[1ex]
&&\left. - 2At'\,\delta_{l1}\,\left<v_j(\bfX, t)\,v_{m2}(\bfX', t')\right>\right]\nonumber\\[1em]
\widehat{\beta}_{il}(\bfx, t, t') &=& \epsilon_{ij2}\left[\left<v_j(\bfX, t)\,v_{1l}(\bfX', t')\right>\right.
\nonumber\\[1ex]
&&\left. - 2At'\,\delta_{l1}\,\left<v_j(\bfX, t)\,v_{12}(\bfX', t')\right>\right]\nonumber\\[1em]
\widehat{\eta}_{iml}(\bfx, t, t') &=& \epsilon_{ijl}\left<v_j(\bfX, t)\,v_m(\bfX', t')\right>
\label{trcoeffsvv}
\end{eqnarray}
\noindent
where the quantities $\bfX = \left(x_1\,,x_2 - 2Atx_1\,,x_3\right)$ and $\bfX' = \left(x_1\,,x_2 - 2At'x_1\,,x_3\right)$. 

We can arrive at some general conclusions for delta--correlated--in--time velocity fields. Let the  
the two--point correlator taken between spacetime 
points $(\bfR, \tau)$ and $(\bfR', \tau')$ be $\left<v_i(\bfR, \tau)\,v_j(\bfR', \tau')\right> 
= \delta(\tau - \tau')\,T_{ij}(\bfR, \bfR', \tau)$. We define $T_{ijl}(\bfR, \tau) = \left(\partial T_{ij}/\partial R'_l\right)_{\bfR'=\bfR}$. The delta--function ensures that $\bfX$ and $\bfX'$  
occuring in the velocity correlators of equation~(\ref{trcoeffsvv})
are equal to each other. So 
$\left<v_i(\bfX,t)\,v_j(\bfX',t')\right> = \delta(t-t')\,T_{ij}(\bfX, \bfX, t)$, and 
$\left<v_i(\bfX,t)\,v_{jl}(\bfX',t')\right> =  \delta(t-t')\,T_{ijl}(\bfX, t)$. The integrals over time in equation~(\ref{emf}) can all be performed, so the mean EMF is
\begin{eqnarray}
{\cal E}_i &=& \epsilon_{ijm}\left[T_{jml} - 2At\,\delta_{l1}\,T_{jm2}\right]
\left[H_l + 2At\,\delta_{l2}\,H_1\right]\nonumber\\  
&-& \epsilon_{ijl}\left[T_{jm} \;+\; 2At\,\delta_{m2}\,T_{j1}\right]H_{lm}
\label{deltaemf}
\end{eqnarray}
\noindent
It is useful to write the EMF in terms of the original variables and lab frame coordinates. To this end we 
transform
\beq
H_{lm} = \left(\frac{\partial}{\partial X_m} - 2A\tau\,\delta_{m1}\,\frac{\partial}{\partial X_2}\right)\,B_l = B_{lm} - 2A\tau\,\delta_{m1}\,B_{l2}
\label{HBgrad}
\eeq
\noindent where $B_{lm} = (\partial B_l/\partial X_m)$. 
Then the explicit dependence of ${\cal E}_i$ 
on the shear parameter $A$ cancels out,
and mean EMF assumes the simple form, 
\beq
{\cal E}_i \;=\; \epsilon_{ijm}\,T_{jml}\,B_l \;-\; \epsilon_{ijl}\,T_{jm}\,B_{lm}
\label{deltaemfsimple}
\eeq
\noindent
which is the familiar expression obtained in the absence of shear. Thus, 
shear needs time to manifest and, to see the effects of shear 
explicitly, it is necessary to consider non zero correlation times. 
Henceforth we consider velocity statistics with finite correlation times.

The linear shear flow has a basic symmetry relating to measurements made by a special subset of 
all observers. We define a comoving observer as one whose velocity with respect to 
the lab frame is equal to the velocity of the background shear flow, and whose Cartesian coordinate
axes are aligned with those of the lab frame. A comoving observer can be labelled by the coordinates, 
$\bfxi = (\xi_1, \xi_2, \xi_3)$, of her origin at time $\tau=0$. Different labels identify different 
comoving observers and vice versa. As the labels run over all possible values, they exhaust the set of 
all comoving observers. The origin of the coordinate axes of a comoving observer translates 
with uniform velocity; its position with respect to the origin of the lab frame is given by 
\beq
\bfX_c(\tau) \;=\; \left(\xi_1\,, \xi_2 - 2A\tau \xi_1\,, \xi_3\right)
\label{orgvector}
\eeq
\noindent
An event with spacetime coordinates $(\bfX, \tau)$ in the lab frame has spacetime coordinates 
$(\tilde{\bfX}, \tilde{\tau})$ with respect to the comoving observer, given by
\beq
\tilde{\bfX} \;=\; \bfX \;-\; \bfX_c(\tau)\,,\qquad \tilde{\tau} \;=\; \tau - \tau_0
\label{coordtr}
\eeq
\noindent 
where the arbitrary constant $\tau_0$ allows for translation in time as well. 

Let $\left[\tilde{\bfB}(\tilde{\bfX}, \tilde{\tau})\,, \tilde{\bfb}(\tilde{\bfX}, \tilde{\tau})\,,\tilde{\bfv}(\tilde{\bfX}, \tilde{\tau})\right]$ denote the mean, the fluctuating magnetic fields and the fluctuating velocity field, respectively, as measured by the comoving observer. They are all equal to the respective quantities measured in the lab frame: $\left[\tilde{\bfB}(\tilde{\bfX},\tilde{\tau})\,,\tilde{\bfb}(\tilde{\bfX},\tilde{\tau})\,,
\tilde{\bfv}(\tilde{\bfX}, \tilde{\tau})\right] = \left[\bfB(\bfX, \tau)\,,\bfb(\bfX, \tau)\,,\bfv(\bfX, \tau)\right]$. That this must be true may be understood as follows. Magnetic fields are invariant under non--relativistic boosts, so the mean and fluctuating magnetic fields must be the same in both frames. To see that the fluctuating velocity fields must also be the same in both frames, we note that the total fluid velocity measured by the comoving observer is, by definition, equal to $\left(-2A\tilde{X}\ey + \tilde{\bfv}(\tilde{\bfX}, \tilde{\tau})\right)$. This must be equal to the difference between the velocity in the lab frame, $\left(-2AX\ey + \bfv(\bfX, \tau)\right)$, and $\left(-2A\xi_1\ey\right)$, which is the velocity of the comoving observer with respect to the lab frame. Using $\tilde{X} = X - \xi_1$, we see that 
$\tilde{\bfv}(\tilde{\bfX}, \tilde{\tau}) = \bfv(\bfX, \tau)$. Equations~(\ref{meanindeqn}) and (\ref{flucindeqn}) are invariant under the simultaneous transformations of spacetime coordinates and fields discussed above. We note that this symmetry property is actually invariance under a subset of the full ten--parameter Galilean group, parametrized by the five quantities $\left(\xi_1, \xi_2, \xi_3, \tau_0, A\right)$; for brevity we refer to this restricted symmetry as Galilean invariance, or simply GI. There is a fundamental difference between the coordinate transformations associated with GI (equation~\ref{coordtr}) and the shearing transformation (equation~\ref{sheartr}). The former relates different comoving observers, whereas the latter describes a time--dependent distortion of the coordinate axes of one observer. Moreover, the relationship between old and new variables is homogeneous for the Galilean transformation, whereas it is  inhomogeneous for the shearing transformation.

Naturally occuring processes lead to G--invariant velocity statistics.
Let the observer in the lab frame correlate $v_i$ at spacetime location $(\bfR, \tau)$ with $v_j$ 
at location $(\bfR', \tau')$. Now consider a comoving observer,
the position vector of whose origin is given by $\bfX_c(\tau)$ 
of equation~(\ref{orgvector}). An identical experiment performed by 
this observer must yield the same results, the measurements now made
at the spacetime points denoted by $(\bfR+\bfX_c(\tau), \tau)$ 
and $(\bfR'+\bfX_c(\tau'), \tau')$ in the lab frame variables.  
Therefore, a GI two--point velocity correlator must satisfy the condition, 
\begin{eqnarray}
&&\left<v_i(\bfR, \tau)\,v_j(\bfR', \tau')\right> =\nonumber\\ 
&&\qquad \left<v_i(\bfR + \bfX_c(\tau), \tau)\,v_j(\bfR' + \bfX_c(\tau'), \tau')\right> 
\label{ginvacorr}
\end{eqnarray}
\noindent
for all $(\bfR, \bfR', \tau, \tau', \bfxi)$.
We also have 
\begin{eqnarray}
&&\left<v_i(\bfR, \tau)\,v_{jl}(\bfR', \tau')\right> =\nonumber\\ 
&&\qquad \left<v_i(\bfR + \bfX_c(\tau), \tau)\,v_{jl}(\bfR' + \bfX_c(\tau'), \tau')\right> 
\label{ginvder}
\end{eqnarray}
\noindent
If we now set $\bfR = \bfR' = {\bf 0}\,, \tau = t\,, \tau'= t'$ and $\bfxi = \bfx$, we will have $\bfX_c(\tau) = \left(x_1\,,x_2 - 2Atx_1\,,x_3\right)$ and $\bfX_c(\tau') = \left(x_1\,,x_2 - 2At'x_1\,,x_3\right)$. Therefore $\bfX_c(\tau)$ and $\bfX_c(\tau')$ are equal to $\bfX$ and $\bfX'$, which are the quantities that enter as arguments in the velocity correlators of equations~(\ref{trcoeffsvv}) defining the transport coefficients. Hence, (reading equations~(\ref{ginvacorr}) and (\ref{ginvder}) from right to left), we see
that
\begin{eqnarray}  
\left<v_i(\bfX,t)\,v_j(\bfX',t')\right> &=&  \left<v_i({\bf 0},t)\,v_j({\bf 0},t')\right> 
= R_{ij}(t, t')\nonumber\\
\left<v_i(\bfX,t)\,v_{jl}(\bfX',t')\right> &=&  \left<v_i({\bf 0},t)\,v_{jl}({\bf 0},t')\right> 
= S_{ijl}(t, t')\nonumber\\
&&\label{rsdef}
\end{eqnarray}
\noindent
are independent of space, and are given by the functions $R_{ij}(t, t')$ and $S_{ijl}(t, t')$. Symmetry
and incompressiblity imply that $R_{ij}(t, t') = R_{ji}(t', t)$ and $S_{ijj}(t, t')=0\,$. Using 
equations~(\ref{rsdef}) in equations~(\ref{trcoeffsvv}), we find that the GI transport coefficients  
\begin{eqnarray}
\widehat{\alpha}_{il}(t, t') &=& \epsilon_{ijm}\left[S_{jml}(t,t') - 2At'\,\delta_{l1}\,S_{jm2}(t,t')\right]\nonumber\\
\widehat{\beta}_{il}(t, t') &=& \epsilon_{ij2}\left[S_{j1l}(t,t') - 2At'\,\delta_{l1}\,S_{j12}(t,t')\right]\nonumber\\
\widehat{\eta}_{iml}(t, t') &=& \epsilon_{ijl}\,R_{jm}(t, t')
\label{trcoeffginv}
\end{eqnarray}
\noindent
are also independent of space. 

Galilean invariance is the fundamental reason that the velocity correlators, hence the transport coefficients, are independent of space. The derivation given above is purely mathematical, relying on the basic freedom of choice of parameters $(\bfR, \bfR', \tau, \tau', \bfxi)$, but we can also understand the results more physically. $\bfX$ and $\bfX'$ can be thought of as the location of the origin of a comoving observer at times $t$ and $t'$, respectively. GI implies that the velocity correlators measured by the comoving observer at her origin at times $t$ and $t'$ must be equal to the velocity correlators measured by any comoving observer at her origin at times $t$ and $t'$. In particular, this must be true for the observer in the lab frame, which explains equations~(\ref{rsdef}), consequently equations~(\ref{trcoeffginv}). We can derive an expression for the GI mean EMF by using equations~(\ref{trcoeffginv}) for the transport coefficients in equation~(\ref{emf}), and simplifying the integrands. Define
\begin{eqnarray}
C_{jml}(t,t') &\;=\;& S_{jml}(t,t') \;-\; 2A(t-t')\delta_{m2}\,S_{j1l}(t,t')\nonumber\\
D_{jm}(t,t') &\;=\;& R_{jm}(t,t') \;+\; 2At'\delta_{m2}\,R_{j1}(t,t')
\label{cddef}
\end{eqnarray}
\noindent
Then the mean EMF, $\bfemf(\bfx, t)$, can be written compactly as
\begin{eqnarray}
{\cal E}_i &=& \epsilon_{ijm}\,\int_0^t dt'\;C_{jml}(t,t')H'_l\nonumber\\
&-& \int_0^t dt'\,\left[\epsilon_{ijl} - 2A(t-t')\delta_{l1}\epsilon_{ij2}\right]\ D_{jm}(t, t')H'_{lm}
\nonumber\\
&&\label{emfcd}
\end{eqnarray}
\noindent

The mean field equation~(\ref{meanindeqn}) for $\bfH(\bfx, t)$ is
\beq
\frac{\partial H_i}{\partial t} + 2A\delta_{i2}H_1 =
\left(\bnabla\cross\bfemf\right)_i + \eta\bnabla^2 H_i
\label{hmeanindeqn}
\eeq
where $\left(\bnabla\right)_p\equiv\partial/\partial X_p = 
\left(\partial/\partial x_p + 2At\delta_{p1}\partial/\partial x_2\right)$. 
We use equation~(\ref{emfcd}) to evaluate $\left(\bnabla\cross\bfemf\right)_i$:

\begin{eqnarray}
\left(\bnabla\cross\bfemf\right)_i &=& \int_0^t dt'\left[C_{iml} - C_{mil}\right]
\left[H'_{lm} + 2At\delta_{m1}H'_{l2}\right]\nonumber\\
&+& \int_0^t dt'\,D_{jm}\left\{ H'_{ijm} + 2At\delta_{j1}H'_{i2m}\right.\nonumber\\ 
&& \left.-2A(t-t')\delta_{i2}\left[H'_{1jm} + 2At\delta_{j1}H'_{12m}\right]\right\}\nonumber\\ 
&&\label{curlemf}
\end{eqnarray}
\noindent
Equations~(\ref{hmeanindeqn}) and (\ref{curlemf}) form a closed set of 
integro--diffential equations governing the dynamics of the mean field, 
$\bfH(\bfx, t)$, valid for arbitrary values of $A$. The most visible properties of 
equation~(\ref{curlemf}) for $\left(\bnabla\cross\bfemf\right)$ 
are: (i) Only the part of $C_{iml}(t,t')$ that is antisymmetric in the indices $(i,m)$ 
contributes. Indeed both $S_{iml}$ and $C_{iml}$ can vanish for non helical velocity 
fluctuations, in which case dynamo action is determined only by the $D_{jm}$ terms. (ii) 
The $D_{jm}(t,t')$ terms are such that $\left(\bnabla\cross\bfemf\right)_i$
involves only $H_i$ for $i=1$ and $i=3$, whereas $\left(\bnabla\cross\bfemf\right)_2$
depends on both  $H_2$ and $H_1$. Together with the mean field induction 
equation~(\ref{hmeanindeqn}) this means that the equations determining the time evolution 
of $H_1$ and $H_3$ are closed. Thus $H_1(\bfx, t)$ (or $H_3(\bfx, t)$) can be computed by
using only the initial data $H_1(\bfx, 0)$ (or $H_3(\bfx, 0)$). The equation for $H_2$ 
involves both $H_2$ and $H_1$, and can then be solved.

The implications for the original field, $\bfB(\bfX, \tau)$, can be read off, because it is 
equal to $\bfH(\bfx, t)$ component--wise (i.e $B_i(\bfX, \tau) = H_i(\bfx, t)$).
Thus, the $D_{jm}(t,t')$ terms do not couple either 
$B_1$ or $B_3$ with any other components, excepting themselves. In demonstrating this, we have 
not assumed that either the shear is small, or that $\bfH(\bfx, t)$ is such a slow function of 
time that it can be pulled out the time integrals in equations~(\ref{emfcd}) and (\ref{curlemf}). 
Comparing with earlier work (where, essentially, both assumptions have been made) we conclude that
{\it there is no shear--current assisted dynamo of the form discussed by \cite{RK03,RS06,RK06}, where 
there is explicit coupling of $B_2$ and $B_1$ in the evolution equation for $B_1$}. 
Our calculations are based on a non perturbative treatment of shear, and this makes for a 
basic departure from earlier work which have treated shear perturbatively. Even 
when the shear is weak, two fluid elements which were close together initially would be separated 
by arbitrarily large distances at late times. Thus the two--time correlators, which appear naturally 
in the dynamo problem, have to be handled carefully in the presence of shear. Moreover, the 
perturbative treatment of shear is not guaranteed to preserve GI, which is a natural and fundamental 
ingredient of our non perturbative approach.

In conclusion we find that systematic use of the shearing coordinate transformation 
and the Galilean invariance of a linear shear flow allows us to develop a quasilinear theory of the 
shear dynamo which, we emphasize, is non perturbative in the shear parameter. Specifically, we have 
proved that there is essentially no shear--current assisted dynamo  in the quasilinear limit 
when FOSA is applicable. Moreoever, our results are valid for any GI velocity statistics, independent of the forces (Coriolis, buoyancy etc) governing the dynamics of the velocity field. However, large--scale 
non helical dynamos (i.e. with no initially imposed kinetic helicity) are not ruled out, and further 
progress requires developing a dynamical theory of velocity correlators in shear flows.

\begin{acknowledgments}
We acknowledge Nordita for providing a stimulating atmosphere
during the program on `Turbulence and Dynamos'.
We thank Axel Brandenburg, Karl-Heinz R\"adler and Matthias Rheinhardt for valuable
comments.
\end{acknowledgments}


\begin{thebibliography}{99}
\bibitem{BRRK08}
A. Brandenburg et al., Astrophys. J., {\bf 676}, 740 (2008).
\bibitem{Yousef08}
T. A. Yousef et al., Phys. Rev. Lett., {\bf 100}, 184501 (2008);
Astron. Nachr. {\bf 329}, 737 (2008). 
\bibitem{PKB} 
P. J. K\"apyl\"a, M. J. Korpi and A. Brandenburg, 
Astron. Astrophys., {\bf 491}, 353 (2008); arXiv:0812.1792 (2008);
D. W. Hughes and M. R. E. Proctor, arXiv:0810.1586 (2008).
\bibitem{RK03}
I. Rogachevskii and N. Kleeorin, Phys. Rev. E {\bf 68}, 036301 
(2003); {\bf 70} 046310 (2004); Astron. Nachr., {\bf 329}, 732 (2008).
\bibitem{Schek08} A. A. Schekocihin et al., arXiv:0810.2225 (2008).
\bibitem{RS06}
K.H. R\"adler and R. Stepanov, Phys. Rev E {\bf 73}, 056311 (2006).
\bibitem{RK06}
G. R\"udiger and L. L. Kitchatinov, Astron. Nachr. {\bf 327}, 298 (2006).
\bibitem{dynam} 
H. K. Moffatt,
Magnetic Field Generation in Electrically Conducting Fluids,
Cambridge University Press, Cambridge (1978);
F. Krause, K.-H. R\"adler,
Mean-field magneto\-hydrodynamics and dynamo theory,
Pergamon Press, Oxford (1980);
A. Brandenburg and K. Subramanian, Phys. Rep. {\bf 417}, 1 (2005).

\end{thebibliography}
\end{document}